\documentclass[review, authoryear]{elsarticle}

\usepackage[utf8]{inputenc} 
\usepackage[T1]{fontenc}    
\usepackage{hyperref}       
\usepackage{url}            
\usepackage{booktabs}       
\usepackage{amsmath}
\usepackage{amssymb}
\usepackage{nicefrac}       
\usepackage{microtype}      
\usepackage{xcolor}         
\usepackage{graphicx}		
\usepackage{multirow}
\usepackage{subcaption}
\usepackage{ulem}
\useunder{\uline}{\ul}{}

\usepackage{comment}

\usepackage{cleveref}
\crefname{enumi}{example}{examples}

\journal{Expert Systems with Applications}

\begin{document}

\begin{frontmatter}

\title{Attention-based sequential recommendation system using multimodal data}

\author[label1]{Hyungtaik Oh}
\ead{colin45@naver.com}
\author[label1]{Wonkeun Jo}
\ead{jowonkun@g.cnu.ac.kr}
\author[label2]{Dongil Kim\corref{cor1}}
\ead{d.kim@ewha.ac.kr}
\affiliation[label1]{organization={Department of Computer Science and Engineering, Chungnam National University},
            addressline={99 Daehak--ro}, 
            city={Yuseong--gu},
            postcode={34134}, 
            state={Daejeon},
            country={Republic of Korea}}
\affiliation[label2]{organization={Department of Data Science, Ewha Womans University},
            addressline={52 Ewhayeodae--gil}, 
            city={Seodaemun--gu},
            postcode={03760}, 
            state={Seoul},
            country={Republic of Korea}}
\cortext[cor1]{Corresponding author.}

\begin{abstract}
Sequential recommendation systems that model dynamic preferences based on a user’s past behavior are crucial to e-commerce. 
Recent studies on these systems have considered various types of information such as images and texts. 
However, multimodal data have not yet been utilized directly to recommend products to users. 
In this study, we propose an attention-based sequential recommendation method that employs multimodal data of items such as images, texts, and categories. 
First, we extract image and text features from pre-trained VGG and BERT and convert categories into multi-labeled forms. 
Subsequently, attention operations are performed independent of the item sequence and multimodal representations. 
Finally, the individual attention information is integrated through an attention fusion function. 
In addition, we apply multitask learning loss for each modality to improve the generalization performance. 
The experimental results obtained from the Amazon datasets show that the proposed method outperforms those of conventional sequential recommendation systems.
\end{abstract}


\end{frontmatter}

\section{Introduction}\label{sec.1}
Recommendation systems provide information on items preferred by users on e-commerce websites \citep{shao2021survey}. 
Although they change dynamically, user preferences can be deduced from the purchase history of users. 
For example, a user who purchases a shirt may subsequently purchase a jacket or pants.
In this manner, a sequential recommendation system models dynamic preferences from the past purchase behavior of users and recommends the next item to purchase \citep{kang2018self}.

\begin{figure}[bp!]
    \centering
    \includegraphics[width=1\textwidth]{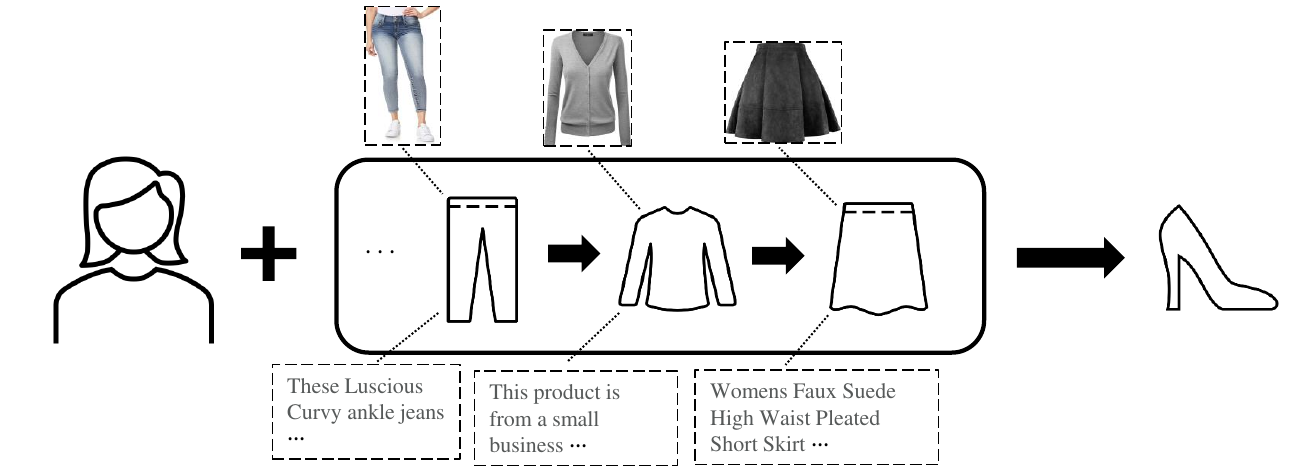}
    \caption{A sequence of items purchased by the user, which contains multimodal data such as images, text, and categories.}
    \label{fig:my_label}
\end{figure}

As shown in Fig. \ref{fig:my_label}, users usually check various sources of information, such as item image, description, and satisfaction scores, on an e-commerce website before purchasing an item. Such multimodal data contain important information about user behavior and can be used to train recommendation models \citep{chen2019pog}. For example, various types of information, such as thumbnails, movie storylines, and casts, can play a key role in video recommendation \citep{wei2019mmgcn}. However, utilizing multimodal data is challenging because they have different forms that require preprocessing steps.

Neural networks have been applied to sequential recommendation systems to improve their performance by utilizing various types of data. 
Convolution neural networks (CNNs) were employed to learn the user sequence patterns as image data, whereas recurrent neural networks (RNNs) were employed to learn the user sequence patterns with recurrent hidden states \citep{hidasi2015session, tang2018personalized}. 
In addition, some studies showed that using an RNN can extract useful information from multimodal data such as images and text \citep{cui2018mv, yan2019merging}. 
Meanwhile, the attention mechanism is a popular calculation module in neural networks and shows promising results when applied to sequential recommendation systems \citep{vaswani2017attention}. 
For example, self-attention networks can solve long-range dependency problems by adaptively assigning weights to the previous items in a sequence. 
Recent research has also attempted to reflect side information, such as categories and positions, in attention-based sequential recommendation systems \citep{xie2022decoupled, zhang2019feature, zhou2020s3}. 
These methods resulted in a significant improvement in performance. 
However, they only used a limited amount of structured information, and not unstructured multimodal data such as images and text.

In this study, we propose a novel attention-based sequential recommendation system using multimodal data. 
Using a \textbf{M}ultimodal \textbf{A}ttention \textbf{F}usion (MAF) method, the proposed system applies multimodal information extracted from images, texts, and categories to predict the next purchase items of users through attention-based sequential recommendation. 
The multimodal data of users’ sequential behavior were collected from the purchase history, item image, item description text, and item category information, and the image and text features were extracted using pre-trained models. 
These data were, subsequently, used as input data for neural networks with attention modules, and attention operations were performed independently of item sequence and multimodal representation. 
The attention fusion function combines each attention result to adjust the different weights of the sequential multimodal representations. 
In the training process, multitask learning was employed to better represent each data modality. 
The experimental results show that the proposed method improved the recommendation performance on the Amazon datasets. 
The main contributions of this study are summarized as follows:

\begin{enumerate}
\item[$\bullet$] The proposed method utilizes multimodal data from images, text, and categories. 
Features are extracted directly from each modality and the attention operations are performed, following which, the information is combined by the MAF.
\item[$\bullet$] Multitask learning is employed for a better generalization performance. 
In the training process, the loss function comprises each task-specific loss for predicting the item ID, image, text, and category.
\item[$\bullet$] Experiments on four datasets collected from Amazon show that our model outperforms other recently proposed models.
\end{enumerate}

\section{Related work}\label{sec.2}

\subsection{Sequential recommendation}\label{sec.2.1}

The recommendation system have been widely studied based on collaborative filtering \citep{sarwar2001item}, matrix factorization \citep{koren2009matrix, mnih2007probabilistic}, top–N recommendation \citep{hu2008collaborative, pan2008one}, and their hybrid method \citep{walek2020hybrid}; however, most of them cannot easily model the sequential purchasing patterns of users. 
To overcome this drawback of the conventional methods, sequential recommendation system has recently been studied to model the purchasing patterns of users. Moreover, deep learning techniques have enabled us to apply sequential recommendation to neural networks to more effectively model user-item interaction. 
For example, the convolutional sequence embedding recommendation model (CASER) \citep{tang2018personalized}, which uses a convolutional filter, considers the item sequence of users as an image and learns sequential information through convolutional operations. 
RNNs have also been utilized with a gated recurrent unit (GRU) to learn the sequential patterns of users (GRU4Rec) \citep{hidasi2015session}. 
The RNN unit enables us to input the hidden state and the current input at each point in time, while considering the sequential dependencies between items purchased by the user. 
In addition, self-attentive sequential recommendation (SASRec) methods perform extremely well owing e to their ability to capture long-range dependencies between items. 
SASRec excels in the unidirectional learning of user sequential patterns through self-attention technique, and computes results faster than previously proposed CNN/RNN-based sequential models \citep{kang2018self}. 
Finally, sequential recommendation with bidirectional encoder representations from transformer (BERT4Rec) is an improved model that performs bidirectional self-attention by applying random masking to the input sequence \citep{sun2019bert4rec}.

\subsection{Attention-based sequential recommendation to reflect side information}\label{sec.2.2}

Previous studies have proved that images and text are useful features for recommending new items \citep{han2021multimodal, he2016vbpr, iqbal2018multimodal, liu2019recommender}. 
Recently, attempts have been made to integrate the side information of items such as category and location into attention-based sequential recommendations. 
SASRecF, which is an improved version of SASRec, uses multimodal item information. 
The embedding layer of SASRecF directly employs the side information of the item with the sequence to be used as the input for the model. 
The feature-level deeper self-attention (FDSA) method integrates several types of side information via a vanilla attention mechanism and performs self-attention operations on the sequences and side information \citep{zhang2019feature}. 
It subsequently concatenates the outputs of sequences and multimodal blocks into fully connected layers for next-item recommendations. 
The decoupled side information fusion for sequential recommendation (DIF-SR) is another method that reflects side information. 
It decouples the information between attention operations and learns the relative importance of each type of side information. 
Consequently, this method has high attention representation capacity and flexibility because the decoupled attention improves the final representation by breaking the rank bottleneck \citep{xie2022decoupled}. 
Moreover, a multi-task training method was used to further activate side formation. 
All the above-mentioned methods consider side information in attention-based sequential recommendation. 
However, the side information does not include complicated information such as images and texts; only simple data such as categories and positions are considered.

\section{Proposed method}
\subsection{Task definition and overall framework}

\begin{figure*}[tb!]
    \centering
    \includegraphics[width=1\textwidth]{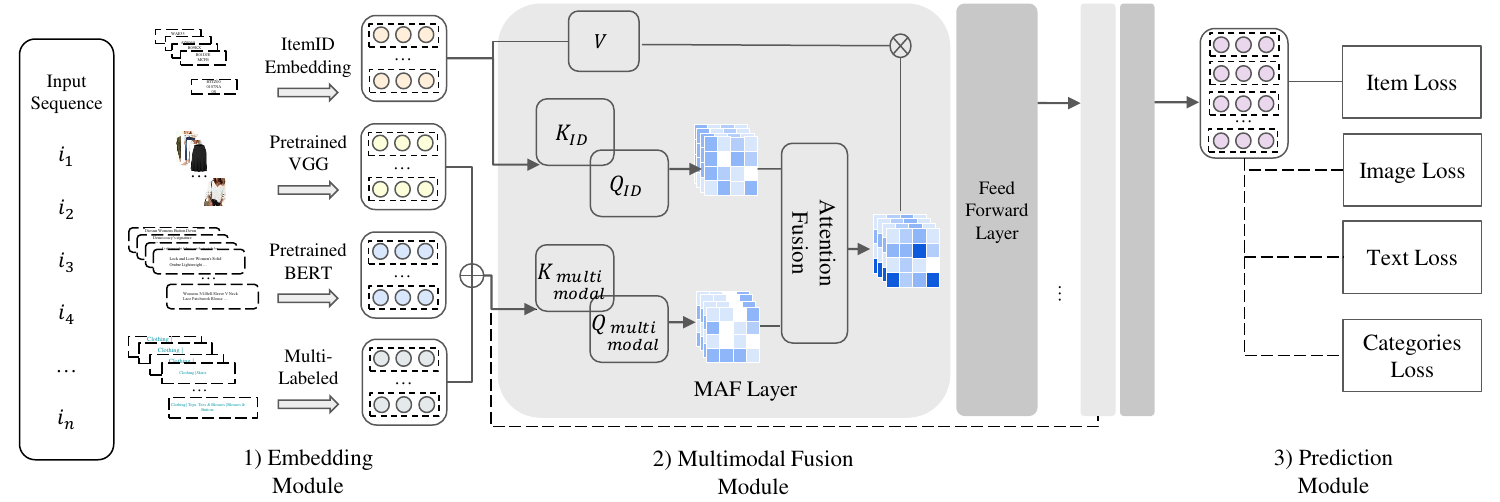}
    \caption{Proposed model’s structure: 1) Features are extracted with pre-trained VGG, BERT, and etc., 2) the attention operation with multimodal representations, and 3) multi-task learning for better generalization.}
    \label{fig:proposal_model}
\end{figure*}

The problem addressed in this study can be formulated with the following notation: Let $I$ and $U$ denote the sets of items and users, respectively. 
For multimodal data, we used $V, T,$ and $C$ to represent sets of images, text, and categories, respectively. 
For each user $u \in U$, we used $i_{1:n} =\{i_1 , i_2 ,\cdots, i_n\}$ to represent the purchased item ID sequence where $n$ represents the sequence length. 
Similarly, $v_{1:n}  = \{v_1, v_2,\cdots, v_n\}, t_{1:n}  = \{t_1, t_2,\cdots, t_n\}$, and $c_{1:n} = \{c_1, c_2,\cdots, c_n\}$ denote the corresponding visual, text and category, respectively. 
We use $i_k$ to represent the items that user $u$ interacted with at the $k$-th time step, and $v_k$, $t_k$, and $c_k$ are the corresponding visuals, texts, and categories of $i_k$. 
Based on these notations, the task of our sequential recommendations is to learn an objective function $f(\cdot)$ that recommends the possible purchase item after the $n$-th item that the user $u$ will interact with at time $i_{n+1}$ given multimodal data from $i_{1:n} \in \mathbb{R}^{n \times d}$, $v_{1:n} \in \mathbb{R}^{n \times d_v}$, $t_{1:n} \in \mathbb{R}^{n \times d_t}$, and $c_{1:n} \in \mathbb{R}^{n \times d_c}$ as shown in Eq. \ref{function}. 
Here, $d$ indicates the latent (hidden) size of item ID, and $d_v, d_t$, and $d_c$ denotes the latent size of each multimodal piece of information. 
The overall model structure is illustrated in Fig. \ref{fig:proposal_model} and comprises three main modules: \textit{Embedding module, Multimodal fusion module, Prediction module}.

\begin{equation}\label{function}
    f(i_{1:n}, v_{1:n}, t_{1:n}, c_{1:n} ) \to  i_{n+1} 
\end{equation}

\subsection{Embedding module}

The embedding module extracts the embedding vector for each data point from the input sequence of the user. 
To use multimodal data, image and text features are extracted using a pre-trained VGG network, denoted as $v_i \in \mathbb{R}^{1 \times d_v}$ and a pre-trained BERT network, denoted as $t_i \in \mathbb{R}^{1 \times d_t}$, respectively. 
Subsequently, the category is transformed into a multi-label form,  denoted as $c_i \in \mathbb{R}^{1 \times d_c}$. 
Because each user has different interaction times with items, the data is pre-processed using a fixed length $n$ for a neural network. 
If the length of the sequence is less than $n$, it replaces the value with padding; otherwise, it is truncated by $n$. 
The user item embeddings $E_{ID}$ and multimodal data embeddings $E_{v}, E_{t}, E_{c}$ are obtained via the embedding module and can be represented by Eq. \ref{embedding}.

\begin{equation}\label{embedding}
\begin{split}
E_{ID}&=\mathrm{concat}([i_1,i_2,\cdots,i_n]) \\ 
E_{v}&=\mathrm{concat}([v_1,v_2,\cdots,v_n]) \\
E_{t}&=\mathrm{concat}([t_1,t_2,\cdots,t_n]) \\ 
E_{c}&=\mathrm{concat}([c_1,c_2,\cdots,c_n])
\end{split}
\end{equation}

\subsection{Multimodal fusion module}

The multimodal fusion module comprises an MAF and feed-forward layer. 
The MAF layer contains independent attention operations and an attention fusion function that combines sequential multimodal representations. 
For the attention operation, multimodal data of different dimensions were processed to obtain data with the same dimension through linear transformations. 
Because images, texts, and categories extracted from each item contains information related to each other, they are added and expressed as one multimodal, as shown in Eq. \ref{image_text_cate_sum}, where the $W_v$, $W_t$, $W_c$, and $d'$ indicate the linear embedding matrix for each data point and the dimension for connecting multimodal data \citep{cui2018mv}.

\begin{equation}\label{image_text_cate_sum}
\begin{split}
E_{m}=E_{v}W_{v}+E_{t}W_{t}+E_{c}W_{c} \\ 
W_{v}\in \mathbb{R}^{d_v \times d'}, W_{t}\in \mathbb{R}^{d_t \times d'}, W_{c}\in \mathbb{R}^{d_c \times d'}
\end{split}
\end{equation}

When  performing the attention operation, the SASRec's attention calculation process was employed \citep{kang2018self}. 
Additionally, multimodal representations are added directly block by block, reducing the computational complexity and preventing overfitting \citep{xie2022decoupled}. 
The formula for the overall structure of the fusion module can be expressed as shown in Eq. \ref{overall structure of the fusion module}: 
$R^k$ denotes the updated sequence that has passed through the $k$--th block from the sequence and the multimodal input, $FFN$ is a fully connected feed-forward network, and $LN$ is the layer normalization. 
When $k = 1$, $R^1$ refers to the $E_{ID}$.

\begin{equation}\label{overall structure of the fusion module}
R^{k+1}=LN(FFN(MAF(R^k, E_{m})))
\end{equation}

In the case of $MAF(\cdot)$, independent attention operations have a flexible gradient and higher rank, thus improving the representation of the model, enabling it to adaptively learn sequences and multimodality \citep{xie2022decoupled}. 
The item ID attention matrix $att_{ID}^i$ is calculated by $W_Q^i, W_K^i, W_V^i \in \mathbb{R}^{d_i \times d_h}$ that is divided by the size $d_h=d/h$ for $h$'s multi heads among all dimensions $d$ from the sequence $R^k$.
Similarly, $W_Q^{m}$, $W_K^{m} \in \mathbb{R}^{d' \times d'_h}$ is used to compute the multimodal attention matrix $att_{m}^i$.
In this case, the multimodal dimension $d'$ should be smaller than or equal to the sequence dimension $d$ to prevent overfitting and to reduce the amount of computation. 
Attention scores for item ID representation and multimodal representation are calculated as follows:

\begin{equation}\label{multimodal Q-K }
\begin{split}
att_{ID}^i&=(R^kW_Q^i)(R^kW_K^i)^\mathsf{T} \\
att_{m}^i&=(E_{m}W_Q^{m})(E_{m}W_K^{m})^\mathsf{T}
\end{split}
\end{equation}

The \textit{Sum}, \textit{Concat}, and \textit{Gate} functions proposed in previous studies were used to combine the sequences and multimodal attention fusion function $F(\cdot)$ \citep{liu2021noninvasive}. 
$MAF_{att}^i$ is obtained with the $W_V^i$ from the attention fusion function $F$ for mixing two attention matrices.
$\sqrt{d}$ is a scale factor that avoids overgrowth values for the inner products. 
Finally, the result outputs of all the attention heads are concatenated and fed into the feed-forward layers and layer normalization. 
This process is represented as follows:
\begin{equation} \label{attention fusion}
\begin{split}
MAF_{att}^i &= F(att_{ID}^i, att_{m}^i) \\ 
MAF_{head}^i &= softmax(\frac{MAF_{att}^i}{\sqrt{d}})(R^kW_V^i) \\
MAF(R^k, E_m) &= \mathrm{concat}([MAF_{head}^1,\cdots, MAF_{head}^i])
\end{split}
\end{equation}

\subsection{Prediction module}
To recommend a new item, the last element $R_n^L \in \mathbb{R}^{1 \times d}$ is taken from the final representation $R^L$. 
$R_n^L$ computes to the inner product with the item embedding table $M_{id} \in \mathbb{R}^{d \times |I|}$, Where $|I|$ represents the total number of distinct items. 
In addition, the softmax function can be used to derive the probability value of the item that the user will purchase.
According to Eq. \ref{item loss}, a cross-entropy loss with softmax helps to decrease the difference in probabilities between the actual user-purchased items and the predicted items. 
\begin{equation}\label{item loss}
\begin{split}
\hat{i}_{n+1}&=softmax(R_n^LM_{id}) \\
L_{id}&=-\sum_{i=1}^{|l|}i_{n+1}\log\hat{i}_{n+1}
\end{split}
\end{equation}

Furthermore, the use of multi-task learning can improve the generalization of the model and obtain more diverse informations from multimodal data \citep{han2021multimodal}. 
For images and text in multimodal data, the reconstruction loss function can be used to compute the difference between the original representation $v_{n+1}, t_{n+1}$ and the representation output from the attention fusion module $\hat{v}_{n+1}, \hat{t}_{n+1}$, as shown in Eq. \ref{reconstruction loss}. 
$W_{v}' \in \mathbb{R}^{d \times d_v }$, $b_{v}' \in \mathbb{R}^{1 \times d_v}$ and $W_{t}' \in \mathbb{R}^{d\times d_t}$, $b_{t}' \in \mathbb{R}^{1 \times d_t}$ are learnable parameters that aim to remove noise generated through layers and enhance the representations of image and text data \citep{chen2012marginalized}. 
$|V|$ and $|T|$ represent the total number of distinct images and the total number of distinct texts.
\begin{equation}\label{reconstruction loss}
\begin{split}
\hat{v}_{n+1} &= R_n^LW'_{v} + b_{v}' \qquad L_{v}=\sum_{v=1}^{|V|}\left\| v_{n+1}-\hat{v}_{n+1}\right\|^2\\
\hat{t}_{n+1} &= R_n^LW'_{t} + b_{t}' \qquad L_{t}=\sum_{t=1}^{|T|}\left\| t_{n+1}-\hat{t}_{n+1}\right\|^2
\end{split}
\end{equation}

For categories that show multiple label forms, the real category $c_{n+1}$ of items and the predicted category $\hat{c}_{n+1}$ from the attention fusion module are learned using a binary cross-entropy loss, as expressed in Eq. \ref{category loss}. $W_{c}' \in \mathbb{R}^{d \times d_c }$ and $b_{c}' \in \mathbb{R}^{1 \times d_c}$ are learnable parameters, $\sigma$ is sigmoid. Where $|C|$ represents the total number of item-specific categories. 
\begin{equation}\label{category loss}
\begin{split}
\hat{c}_{n+1} &= \sigma(R_n^LW_{c}' + b_{c}')\\
L_{c}&=-\sum_{c=1}^{|C|}c_{n+1}\log\hat{c}_{n+1}+(1-c_{n+1})\log(1-\hat{c}_{n+1})
\end{split}
\end{equation}
The final loss function comprises the sum of the sequence loss and the multimodal loss, as in Eq. \ref{loss total}. $\lambda$ is a hyperparameter that adjusts for the influence of each loss function.
\begin{equation}\label{loss total}
L=L_{id}+\lambda \sum_{j\in{v, t, c}}L_{j}
\end{equation}

\section{Experiments}
\subsection{Datasets and evaluation metrics}

We used the Video Games and Clothing\&Shoes datasets provided by Amazon to evaluate the proposed model. 
The Clothing\&Shoes dataset contained three categories: Boys\&Girls, Women, and Men. 
To consider user trends, we selected only user transaction data from 2016 to 2017. 
We removed items that did not have multimodal features such as images, descriptions, and categories. 
In addition, users and items with fewer than 5 sequential interactions were excluded.
To use multimodal data, we utilized the pre-trained VGG and BERT to obtain 4096 and 768 dimensional representations for image and text, respectively, and categorical data were converted into a multi-label form. 
Table \ref{Statistics of datasets for experiments} summarizes the dataset used in the experiment. 
In an experimental evaluation method, the proposed model was evaluated using a leave-one-out strategy as in previous studies \citep{kang2018self, sun2019bert4rec}. 
The last item in the user-purchased sequence is used as the test data, and the item before the last one is used as the validation data. 
We used Recall@k and NDCG@k as the evaluation metrics as shown in Eq. \ref{measures}. 
Recall@k indicates whether the user includes recommended item k among the items of interest. 
NDCG@k applies weight values differently in the order of preference, assigning higher weights to higher orders ($G_i$ and $G_i^{ideal}$ denote a relevant item and an ideal sequence of relevant items, respectively).
In our experiments, the value of k was set to 10 and 20 to compare the results of Recall@k and NDCG@k.

\begin{table}[tb]
\caption{Statistics of datasets for experiments.}
\label{Statistics of datasets for experiments}
\resizebox{\columnwidth}{!}{%
\begin{tabular}{lcccc}
\hline
Dataset     &  Video Games &  Boys\&Girls & Women & Men \\ \hline
\#Users      & 4,940 & 1,405 & 105,558 & 51,892 \\
\#Items      & 11,002 & 15,603 & 122,701 & 56,325 \\
\#Avg. Actions/User      & 7.4 & 6.8 & 7.7 & 7.4 \\
\#Avg. Actions/Item      & 16.3 & 35 & 22.4 & 35.8 \\
\#Interactions        & 36,752 & 9,587 & 810,245 & 383,148 \\ \hline
\end{tabular}
}
\end{table}

\begin{equation}\label{measures}
\begin{split}
Recall@k = \frac{\# \mathrm{~of~recommendations~that~are~relevant}}{\# \mathrm{~of~all~possible~relevant~items}}\\
NDCG@k = \frac{\mathrm{DCG@k}}{\mathrm{IDCG@k}}\\
where~ DCG@k = \sum_{i=1}^{k}\frac{G_i}{log_2(i+1)}, IDCG@k = \sum_{i=1}^{k_{ideal}}\frac{G_i^{ideal}}{log_2(i+1)}
\end{split}
\end{equation}

\subsection{Hyperparameter settings}
The proposed and baseline models used the RecBole \citep{zhao2021recbole} framework and summarized the results with five random seed changes.
The hyperparameters of the experimental model were trained with a learning rate of 1e-4 and 200 epochs using the Adam optimizer. 
The hidden size $d$ was set to 256 for the proposed as well as all the attention-based models.
Similarly, the attention layer $L$ was set to 4, and the attention head $h$ was set to 8. 
Other hyperparameters use grid search to find the best settings for the model. 
The searching space was: hidden size for multimodal $d'$ $\in$  \{32, 64, 128, 256\}, \textit{$\lambda$} $\in$  \{5, 10, 15, 20, 25\}, $F$ $\in$  \{\textit{Gate}, \textit{Concat}, and \textit{Sum}\}.

\subsection{Benchmark methods}
The model shown below is the baseline models that were used for the experiments. There are two types of models: basic sequential recommendation methods that use only item sequences (\ref{GRU4}--\ref{bert4recc}) and sequential recommendation methods that use multimodal information (\ref{sasrecf} and \ref{fdsa}). 
The hyperparameters of the baseline model used a default setting and the attention-based model used the same layer, head and hidden sizes as the proposed model. 

\begin{enumerate}
    \item\label{GRU4} GRU4Rec \citep{hidasi2015session}: A sequential recommendation model using RNNs. 
    \item\label{caser} Caser \citep{tang2018personalized}: A CNN-based sequential recommendation model that uses horizontal and vertical convolution filters.
    \item\label{sasrec} SASRec \citep{kang2018self}: An attention-based model that learns sequential patterns using self-attention.
    \item\label{bert4recc} BERT4Rec \citep{sun2019bert4rec}: A bidirectional attention-based model that masks a user's sequential patterns for learning.
    \item\label{sasrecf} SASRecF: An extension of the SASRec model that fuses items and their corresponding side information into embedding layers.
    \item\label{fdsa} FDSA \citep{zhang2019feature}: A self-attention model that learns the sequential patterns and side information of users through independent self-attention blocks.
\end{enumerate}

\begin{table*}[th]
\centering
\caption{Performance comparison between the benchmark model and the proposed model.}
\label{tab:performance}
\resizebox{\textwidth}{!}{
\begin{tabular}{c|c|cccccc|c}
\hline
Dataset                                                                 & Metric                         & GRU4Rec & Caser  & SASRec       & BERT4Rec & SASRecF         & FDSA            & Proposal        \\ \hline
\multirow{4}{*}{\begin{tabular}[c]{@{}l@{}}Video\\ Games\end{tabular}} & Recall@10                      & 0.1120  & 0.0733 & {\ul 0.1689} & 0.0705   & 0.1513          & 0.1518          & \textbf{0.1714} \\
                                                                        & Recall@20                      & 0.1733  & 0.1136 & {\ul 0.2429} & 0.1263   & 0.2154          & 0.2120          & \textbf{0.2499} \\
                                                                        & NDCG@10                        & 0.0615  & 0.0406 & 0.0904       & 0.0344   & 0.0907          & \textbf{0.0959} & {\ul 0.0941}    \\
                                                                        & NDCG@20                        & 0.0769  & 0.0507 & 0.1090       & 0.0483   & 0.1069          & {\ul 0.1111}    & \textbf{0.1138} \\ \hline
\multirow{4}{*}{Boys\&Girls}                                            & Recall@10                      & 0.6859  & 0.7372 & {\ul 0.7877} & 0.6774   & 0.7791          & 0.7688          & \textbf{0.7906} \\
                                                                        & Recall@20                      & 0.7123  & 0.7675 & 0.8056       & 0.7251   & {\ul 0.8107}    & 0.7893          & \textbf{0.8262} \\
                                                                        & NDCG@10                        & 0.6430  & 0.6702 & 0.7112       & 0.3292   & \textbf{0.7151} & 0.7092          & {\ul 0.7125}    \\
                                                                        & NDCG@20                        & 0.6496  & 0.6774 & 0.7157       & 0.3412   & \textbf{0.7230} & 0.7144          & {\ul 0.7216}    \\ \hline
\multirow{4}{*}{Women}                                                  & \multicolumn{1}{l|}{Recall@10} & 0.2048  & 0.1817 & {\ul 0.2323} & 0.1806   & 0.2283          & 0.2306          & \textbf{0.2325} \\
                                                                        & \multicolumn{1}{l|}{Recall@20} & 0.2109  & 0.1889 & {\ul 0.2419} & 0.1926   & 0.2370          & 0.2395          & \textbf{0.2422} \\
                                                                        & \multicolumn{1}{l|}{NDCG@10}   & 0.1932  & 0.1661 & 0.2162       & 0.0927   & 0.2151          & \textbf{0.2171} & \textbf{0.2171} \\
                                                                        & \multicolumn{1}{l|}{NDCG@20}   & 0.1948  & 0.1679 & 0.2186       & 0.0957   & 0.2173          & {\ul 0.2193}    & \textbf{0.2196} \\ \hline
\multirow{4}{*}{Men}                                                    & \multicolumn{1}{l|}{Recall@10} & 0.4561  & 0.4285 & 0.4857       & 0.4193   & 0.4863          & {\ul 0.4865}    & \textbf{0.4884} \\
                                                                        & \multicolumn{1}{l|}{Recall@20} & 0.4659  & 0.4398 & {\ul 0.4978} & 0.4354   & 0.4965          & 0.4966          & \textbf{0.5001} \\
                                                                        & \multicolumn{1}{l|}{NDCG@10}   & 0.4390  & 0.4060 & 0.4661       & 0.2291   & 0.4658          & {\ul 0.4677}    & \textbf{0.4691} \\
                                                                        & \multicolumn{1}{l|}{NDCG@20}   & 0.4414  & 0.4088 & 0.4692       & 0.2332   & 0.4684          & {\ul 0.4703}    & \textbf{0.4720} \\ \hline
\end{tabular}%
}
\end{table*}

\subsection{Experimental results}

Table \ref{tab:performance} compares the proposed model with various sequential recommendations on four datasets collected from Amazon. 
\textbf{Bold} text in the table indicates the best result of the comparative models, and \underline{underlined} text denotes the second-best result. 
First, the result for the GRU4Rec, Caser, SASRec, and BERT4Rec models, which cannot receive multimodal information, show that the attention-based SASRec perform best, which demonstrates that the attention mechanism is often a useful structure for learning sequential patterns fo users. 
Attention-based sequential recommendations SASRecF and FDSA that can receive multimodal information often perform worse than SASRec which uses only user-item information. 
In the case of SASRecF, which integrates multimodal data via an embedding layer, the model receives diverse information and has difficulty learning sequential patterns of the users. 
In contrast, the FDSA model learns user sequence patterns and multimodal data via an independent self-attention block. 
Neverthelss, FDSA consistently performs worse when it used complex multimodal data than when it used only user sequences. 
In all the datasets, the proposed model mostly outperforms the baselines in terms of the metrics.
Furthermore, various types of data contribute to different performance improvements depending on the dataset size. 
For large-scale datasets such as the Women and Men datasets, the use of Recall@20 improves the performance of the proposed model by 0.12\% and 0.46\%, respectively, compared with the SASRec model, which only considers sequences. 
For small-scale datasets, such as Video Games and Boys \& Girls the corresponding performance increased by 2.9\% and 2.6\%, respectively, indicating that multimodal data has a significant effect on relatively small datasets.

\subsection{Analysis on attention fusion functions}

Table \ref{Attention fusion function comparison} compares the overall performance when applying different attention fusion functions to the four experimental datasets. 
\textbf{Bold} in the table indicates the best result among the models. Three types of attention fusion, viz. 
\textit{Gate}, \textit{Concat}, and \textit{Sum}, were used to aggregate sequence representation and multimodal representation after performing query-key attention operations independently. 
The results in Table \ref{Attention fusion function comparison} show small performance differences between attention fusions. 
However, \textit{Concat} and \textit{Gate} perform poorly on most of the datasets.
the simple attention fusion function \textit{Sum} performs best on the four datasets without increasing the capacity of the model.

\begin{table}[tb]
\centering
\caption{Attention fusion function comparison experiment result.}
\label{Attention fusion function comparison}
{\small
\begin{tabular}{l|l|lll}
\hline
Dataset                                                                  & Metric    & \textit{Sum}   & \textit{Concat} & \textit{Gate}  \\ \hline
\multirow{4}{*}{\begin{tabular}[c]{@{}l@{}}Video\\ Games\end{tabular}}   & Recall@10 & 0.1714          & \textbf{0.1729}  & 0.1725          \\
                                                                         & Recall@20 & \textbf{0.2499} & 0.2479           & 0.2469          \\
                                                                         & NDCG@10   & \textbf{0.0941} & 0.0929           & 0.0937          \\
                                                                         & NDCG@20   & \textbf{0.1138} & 0.1117           & 0.1124          \\ \hline
\multirow{4}{*}{\begin{tabular}[c]{@{}l@{}}Boys \&\\ Girls\end{tabular}} & Recall@10 & 0.7883 & 0.7855           & \textbf{0.7910}          \\
                                                                         & Recall@20 & \textbf{0.8208} & 0.8180           & 0.8187          \\
                                                                         & NDCG@10   & 0.7099          & 0.7074           & \textbf{0.7134} \\
                                                                         & NDCG@20   & 0.7182          & 0.7156           & \textbf{0.7204} \\ \hline
\multirow{4}{*}{Women}                                                   & Recall@10 & \textbf{0.2327} & 0.2314           & 0.2319          \\
                                                                         & Recall@20 & \textbf{0.2421} & 0.2410           & 0.2414          \\
                                                                         & NDCG@10   & \textbf{0.2163} & 0.2153           & 0.2151          \\
                                                                         & NDCG@20   & \textbf{0.2186} & 0.2177           & 0.2175          \\ \hline
\multirow{4}{*}{Men}                                                     & Recall@10 & \textbf{0.4880} & 0.4867           & 0.4875          \\
                                                                         & Recall@20 & \textbf{0.4995} & 0.4978           & 0.4987          \\
                                                                         & NDCG@10   & \textbf{0.4685} & 0.4682           & 0.4655          \\
                                                                         & NDCG@20   & \textbf{0.4714} & 0.4710           & 0.4683          \\ \hline
\end{tabular}%
}
\end{table}

\subsection{Analysis on utilizing multimodal data}

\begin{figure*}[!th]
    \centering
    \includegraphics[width=1\textwidth]{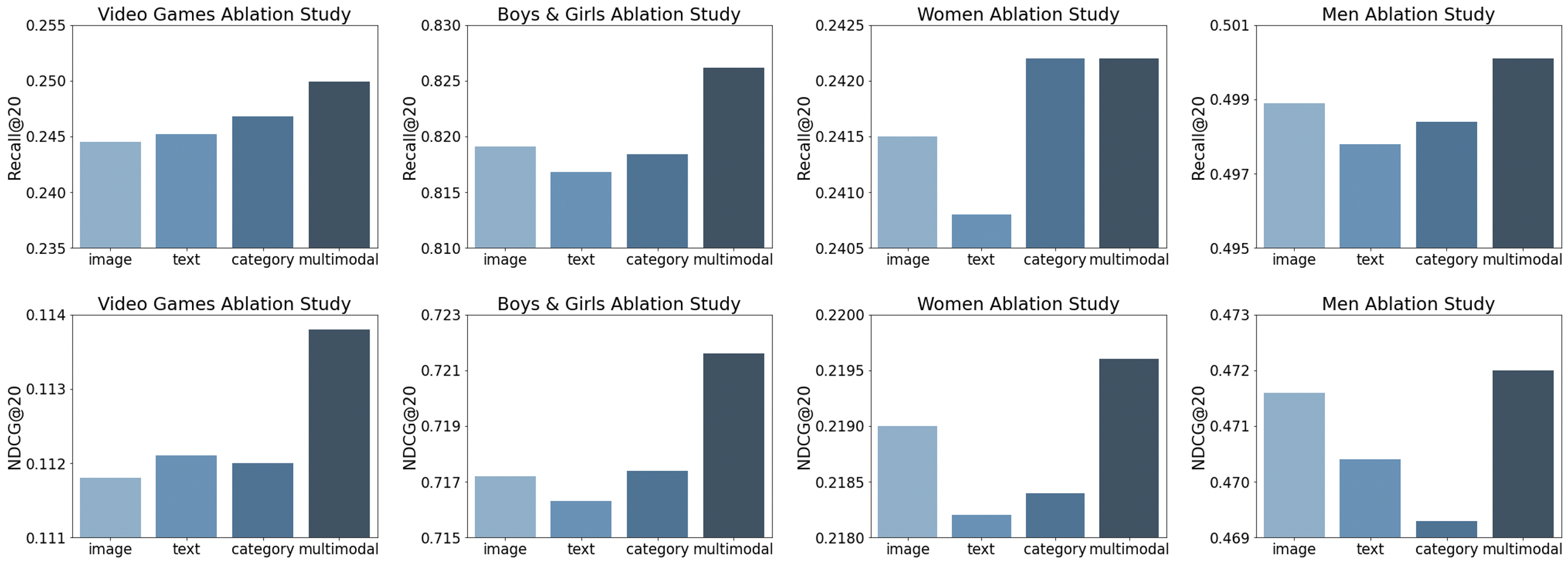}
    \caption{Comparison of multimodal data performance of four datasets.}
    \label{fig:comparison of multimodal data}
\end{figure*}

Fig. \ref{fig:comparison of multimodal data} shows a comparative experiment using images, text, and categories to determine the contribution of various multimodal data in the proposed model. 
The results show that each dataset and evaluation metric yielded different results. 
In Recall@20, item categories of the video game and women datasets perform best among the multimodal data.
For Boys \& Girls and Men, images containing visual information about clothing perform the best.
In the case of NDCG@20, the image data show the best results for the Women and Men datasets. 
In addition, textual data performs best on the Video Games dataset, and category data performs best on the Boys \& Girls dataset. 
Both results indicate that multimodal data have different levels of importance depending on the characteristics of the dataset.
However, on combining images, text, and categories to form multimodal data, they complement each other, resulting in improved model performance on most experimental datasets.

\subsection{Analysis on multi-task loss}
To examine the effect of the multitask learning method applied to the proposed model, comparative experiments were performed with and without multitask loss, as shown in Table \ref{MTL comparison}. 
In most cases, the use of multitask loss resulted in better model performance because it can activate beneficial interactions of multimodal data.

\begin{table}[!t]
\centering
\caption{Performance comparison with and without multi-task learning (MTL) applied to the proposed model: $\%$ Improved indicates the percentage of performance improved with MTL.}
\label{MTL comparison}
{\small
\begin{tabular}{c|c|ccc}
\hline
Dataset                                                                 & Metric                         & w/ MTL & w/o MTL & $\%$ Improved \\ \hline
\multirow{4}{*}{\begin{tabular}[c]{@{}c@{}}Video \\ Games\end{tabular}} & Recall@10                      & 0.1714             & 0.1732                & -1.05\% \\
                                                                        & Recall@20                      & 0.2499             & 0.2465                & 1.38\%  \\
                                                                        & NDCG@10                        & 0.0941             & 0.0934                & 0.75\%  \\
                                                                        & NDCG@20                        & 0.1138             & 0.1119                & 1.7\%   \\ \hline
\multirow{4}{*}{Boys\&Girls}                                            & Recall@10                      & 0.7906             & 0.7926                & -0.25\% \\
                                                                        & Recall@20                      & 0.8262             & 0.8225                & 0.45\%  \\
                                                                        & NDCG@10                        & 0.7125             & 0.7123                & 0.03\%  \\
                                                                        & NDCG@20                        & 0.7216             & 0.7199                & 0.24\%    \\ \hline
\multirow{4}{*}{Women}                                                  & \multicolumn{1}{l|}{Recall@10} & 0.2325             & 0.2321                & 0.17\%  \\
                                                                        & \multicolumn{1}{l|}{Recall@20} & 0.2422             & 0.2410                & 0.5\%   \\
                                                                        & \multicolumn{1}{l|}{NDCG@10}   & 0.2171             & 0.2157                & 0.65\%  \\
                                                                        & \multicolumn{1}{l|}{NDCG@20}   & 0.2196             & 0.2179                & 0.78\%  \\ \hline
\multirow{4}{*}{Men}                                                    & \multicolumn{1}{l|}{Recall@10} & 0.4884             & 0.4873                & 0.23\%  \\
                                                                        & \multicolumn{1}{l|}{Recall@20} & 0.5001             & 0.4985                & 0.32\%  \\
                                                                        & \multicolumn{1}{l|}{NDCG@10}   & 0.4691             & 0.4680                & 0.24\%  \\
                                                                        & \multicolumn{1}{l|}{NDCG@20}   & 0.4720             & 0.4709                & 0.23\%  \\ \hline
\end{tabular}%
}
\end{table}

\subsection{Attention weight visualization}

\begin{figure}[tb]
    \centering
    \includegraphics[width=1\columnwidth]{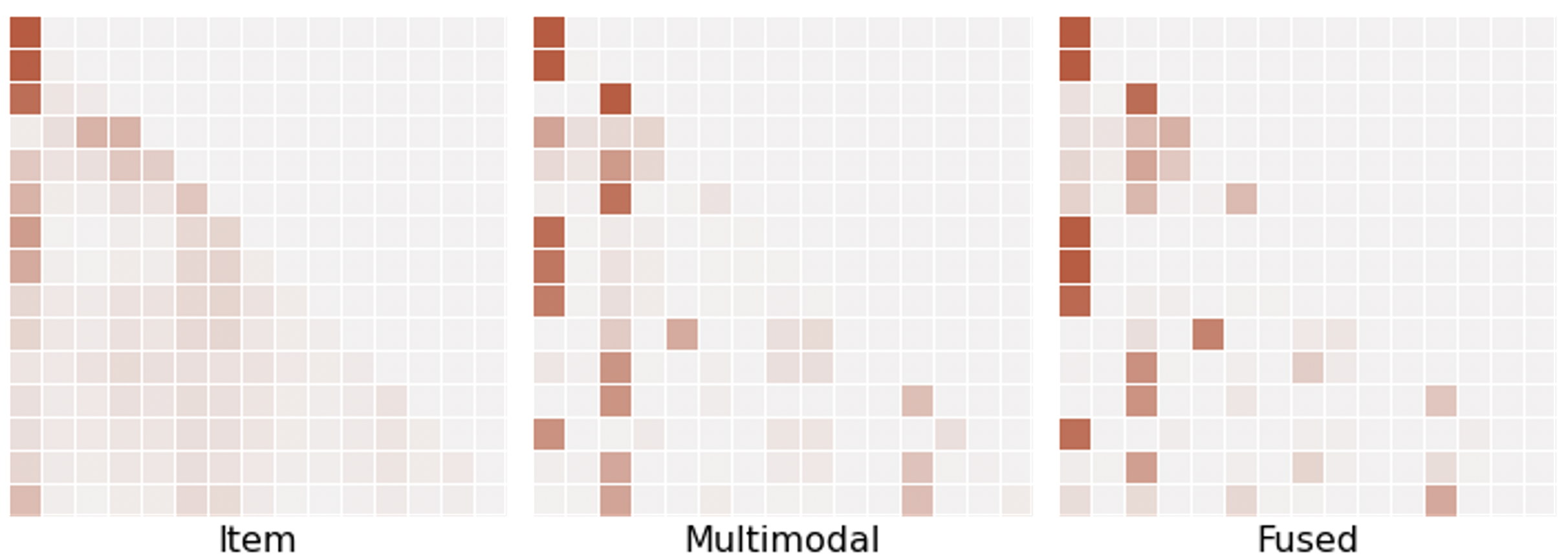}
    \caption{Video Games attention weight visualization.}
    \label{fig:attention weight}
\end{figure}

Fig. \ref{fig:attention weight}. shows the attention weight, which provides an interpretation of the model. 
The Figure shows the results visualized through the last attention layer of the video game dataset extracted from the randomly sampled attention heads.
Because the MAF focused on the interactions between the past purchased items and multimodal features, we applied the masked attention operation.
The last $15^{th}$ time and position of the length of the entire sequence are represented in the Figure; a darker color indicates a higher attention weight. 
The item attention matrix and multimodal attention matrix represent the different attention weights, which demonstrates that the sequence and multimodal data have different impacts when the model performs operations.
In addition, the fusion attention matrix coordinates the contributions of sequence and multimodal data and integrates them into significant patterns.

\section{Conclusion}

In this study, we proposed an attention-based sequential recommendation system model that utilizes multimodal data such as images, texts, and categories. 
The image and text features of the items were extracted from pre-trained VGG and BERT, and the category was converted into a multi-label form to add multimodal information. 
In the proposed method, as each feature information is combined with one multimodal representation, the attention operation is performed by separating the item and multimodal representations. 
In addition, through multi-task learning, both item representation and multimodal representation were learned to refine multimodal information effectively. 
The experiments were conducted on a total of four datasets: Video Games, Boys\&Girls, Women, and Men, which were collected from Amazon. 
The use of the multimodal information in the sequential recommendation system results in a significant performance improvement. 
Through experiments and attention weight visualization, the MAF of the proposed method aids in complementing the ID attention with multimodal information.

However, the applicability of the proposed method is limited by its relatively high computational cost. 
Therefore, in future work, we plan to improve the efficiency of the proposed method using a compressed attention operation to implement an online real-time recommendation system. 
Furthermore, in this study, experiments were conducted on only four categories of the Amazon dataset. 
We plan to expand our experiments to include other datasets.

\section*{CRediT authorship contribution statement}
\textbf{Hyuyngtaik Oh}: Conceptualization, Software, Validation, Writing -- Original Draft. \textbf{Wonkeun Jo}: Methodology, Software, Writing -- Original Draft.  \textbf{Dongil Kim}: Conceptualization, Methodology, Supervision, Writing -- Review \& Editing.

\section*{Declaration of Competing Interest}
The authors declare that they have no known competing financial interests or personal relationships that could have appeared to influence the work reported in this paper.

\section*{Acknowledgments}
This work was supported by the Ewha Womans University Research Grant of 2023. This work was partially supported by Institute of Information $\&$ communications Technology Planning $\&$ Evaluation (IITP) grant funded by the Korea government(MSIT) (No. RS-2022-00155966, Artificial Intelligence Convergence Innovation Human Resources Development (Ewha Womans University)), Research-centric Hospital R$\&$D grant funded by the Korean government (MOHW) (No. HI21C0198), and Basic Science Research Program through the National Research Foundation of Korea(NRF) funded by the Ministry of Education(No. 2022R1A6A3A13056750).

\bibliography{reference}

\begin{thebibliography}{26}
\expandafter\ifx\csname natexlab\endcsname\relax\def\natexlab#1{#1}\fi
\providecommand{\url}[1]{\texttt{#1}}
\providecommand{\href}[2]{#2}
\providecommand{\path}[1]{#1}
\providecommand{\DOIprefix}{doi:}
\providecommand{\ArXivprefix}{arXiv:}
\providecommand{\URLprefix}{URL: }
\providecommand{\Pubmedprefix}{pmid:}
\providecommand{\doi}[1]{\href{http://dx.doi.org/#1}{\path{#1}}}
\providecommand{\Pubmed}[1]{\href{pmid:#1}{\path{#1}}}
\providecommand{\bibinfo}[2]{#2}
\ifx\xfnm\relax \def\xfnm[#1]{\unskip,\space#1}\fi
\bibitem[{Chen et~al.(2012)Chen, Xu, Weinberger and Sha}]{chen2012marginalized}
\bibinfo{author}{Chen, M.}, \bibinfo{author}{Xu, Z.},
  \bibinfo{author}{Weinberger, K.}, \bibinfo{author}{Sha, F.},
  \bibinfo{year}{2012}.
\newblock \bibinfo{title}{Marginalized denoising autoencoders for domain
  adaptation}.
\newblock \bibinfo{journal}{arXiv preprint arXiv:1206.4683} .
\bibitem[{Chen et~al.(2019)Chen, Huang, Xu, Guo, Guo, Sun, Li, Pfadler, Zhao
  and Zhao}]{chen2019pog}
\bibinfo{author}{Chen, W.}, \bibinfo{author}{Huang, P.}, \bibinfo{author}{Xu,
  J.}, \bibinfo{author}{Guo, X.}, \bibinfo{author}{Guo, C.},
  \bibinfo{author}{Sun, F.}, \bibinfo{author}{Li, C.},
  \bibinfo{author}{Pfadler, A.}, \bibinfo{author}{Zhao, H.},
  \bibinfo{author}{Zhao, B.}, \bibinfo{year}{2019}.
\newblock \bibinfo{title}{Pog: personalized outfit generation for fashion
  recommendation at alibaba ifashion}, in: \bibinfo{booktitle}{Proceedings of
  the 25th ACM SIGKDD international conference on knowledge discovery \& data
  mining}, pp. \bibinfo{pages}{2662--2670}.
\bibitem[{Cui et~al.(2018)Cui, Wu, Liu, Zhong and Wang}]{cui2018mv}
\bibinfo{author}{Cui, Q.}, \bibinfo{author}{Wu, S.}, \bibinfo{author}{Liu, Q.},
  \bibinfo{author}{Zhong, W.}, \bibinfo{author}{Wang, L.},
  \bibinfo{year}{2018}.
\newblock \bibinfo{title}{Mv-rnn: A multi-view recurrent neural network for
  sequential recommendation}.
\newblock \bibinfo{journal}{IEEE Transactions on Knowledge and Data
  Engineering} \bibinfo{volume}{32}, \bibinfo{pages}{317--331}.
\bibitem[{Han et~al.(2021)Han, Niu and Wang}]{han2021multimodal}
\bibinfo{author}{Han, T.}, \bibinfo{author}{Niu, S.}, \bibinfo{author}{Wang,
  P.}, \bibinfo{year}{2021}.
\newblock \bibinfo{title}{Multimodal-adaptive hierarchical network for
  multimedia sequential recommendation}.
\newblock \bibinfo{journal}{Pattern Recognition Letters} \bibinfo{volume}{152},
  \bibinfo{pages}{10--17}.
\bibitem[{He and McAuley(2016)}]{he2016vbpr}
\bibinfo{author}{He, R.}, \bibinfo{author}{McAuley, J.}, \bibinfo{year}{2016}.
\newblock \bibinfo{title}{Vbpr: visual bayesian personalized ranking from
  implicit feedback}, in: \bibinfo{booktitle}{Proceedings of the AAAI
  conference on artificial intelligence}.
\bibitem[{Hidasi et~al.(2015)Hidasi, Karatzoglou, Baltrunas and
  Tikk}]{hidasi2015session}
\bibinfo{author}{Hidasi, B.}, \bibinfo{author}{Karatzoglou, A.},
  \bibinfo{author}{Baltrunas, L.}, \bibinfo{author}{Tikk, D.},
  \bibinfo{year}{2015}.
\newblock \bibinfo{title}{Session-based recommendations with recurrent neural
  networks}.
\newblock \bibinfo{journal}{arXiv preprint arXiv:1511.06939} .
\bibitem[{Hu et~al.(2008)Hu, Koren and Volinsky}]{hu2008collaborative}
\bibinfo{author}{Hu, Y.}, \bibinfo{author}{Koren, Y.},
  \bibinfo{author}{Volinsky, C.}, \bibinfo{year}{2008}.
\newblock \bibinfo{title}{Collaborative filtering for implicit feedback
  datasets}, in: \bibinfo{booktitle}{2008 Eighth IEEE international conference
  on data mining}, \bibinfo{organization}{Ieee}. pp. \bibinfo{pages}{263--272}.
\bibitem[{Iqbal et~al.(2018)Iqbal, Kovac and Aryafar}]{iqbal2018multimodal}
\bibinfo{author}{Iqbal, M.}, \bibinfo{author}{Kovac, A.},
  \bibinfo{author}{Aryafar, K.}, \bibinfo{year}{2018}.
\newblock \bibinfo{title}{A multimodal recommender system for large-scale
  assortment generation in e-commerce}.
\newblock \bibinfo{journal}{arXiv preprint arXiv:1806.11226} .
\bibitem[{Kang and McAuley(2018)}]{kang2018self}
\bibinfo{author}{Kang, W.C.}, \bibinfo{author}{McAuley, J.},
  \bibinfo{year}{2018}.
\newblock \bibinfo{title}{Self-attentive sequential recommendation}, in:
  \bibinfo{booktitle}{2018 IEEE international conference on data mining
  (ICDM)}, \bibinfo{organization}{IEEE}. pp. \bibinfo{pages}{197--206}.
\bibitem[{Koren et~al.(2009)Koren, Bell and Volinsky}]{koren2009matrix}
\bibinfo{author}{Koren, Y.}, \bibinfo{author}{Bell, R.},
  \bibinfo{author}{Volinsky, C.}, \bibinfo{year}{2009}.
\newblock \bibinfo{title}{Matrix factorization techniques for recommender
  systems}.
\newblock \bibinfo{journal}{Computer} \bibinfo{volume}{42},
  \bibinfo{pages}{30--37}.
\bibitem[{Liu et~al.(2021)Liu, Li, Cai, Dong, Zhu and
  Shang}]{liu2021noninvasive}
\bibinfo{author}{Liu, C.}, \bibinfo{author}{Li, X.}, \bibinfo{author}{Cai, G.},
  \bibinfo{author}{Dong, Z.}, \bibinfo{author}{Zhu, H.},
  \bibinfo{author}{Shang, L.}, \bibinfo{year}{2021}.
\newblock \bibinfo{title}{Noninvasive self-attention for side information
  fusion in sequential recommendation}, in: \bibinfo{booktitle}{Proceedings of
  the AAAI Conference on Artificial Intelligence}, pp.
  \bibinfo{pages}{4249--4256}.
\bibitem[{Liu et~al.(2019)Liu, Wang, Tang, Yang, Huang and
  Liu}]{liu2019recommender}
\bibinfo{author}{Liu, T.}, \bibinfo{author}{Wang, Z.}, \bibinfo{author}{Tang,
  J.}, \bibinfo{author}{Yang, S.}, \bibinfo{author}{Huang, G.Y.},
  \bibinfo{author}{Liu, Z.}, \bibinfo{year}{2019}.
\newblock \bibinfo{title}{Recommender systems with heterogeneous side
  information}, in: \bibinfo{booktitle}{The world wide web conference}, pp.
  \bibinfo{pages}{3027--3033}.
\bibitem[{Mnih and Salakhutdinov(2007)}]{mnih2007probabilistic}
\bibinfo{author}{Mnih, A.}, \bibinfo{author}{Salakhutdinov, R.R.},
  \bibinfo{year}{2007}.
\newblock \bibinfo{title}{Probabilistic matrix factorization}.
\newblock \bibinfo{journal}{Advances in neural information processing systems}
  \bibinfo{volume}{20}.
\bibitem[{Pan et~al.(2008)Pan, Zhou, Cao, Liu, Lukose, Scholz and
  Yang}]{pan2008one}
\bibinfo{author}{Pan, R.}, \bibinfo{author}{Zhou, Y.}, \bibinfo{author}{Cao,
  B.}, \bibinfo{author}{Liu, N.N.}, \bibinfo{author}{Lukose, R.},
  \bibinfo{author}{Scholz, M.}, \bibinfo{author}{Yang, Q.},
  \bibinfo{year}{2008}.
\newblock \bibinfo{title}{One-class collaborative filtering}, in:
  \bibinfo{booktitle}{2008 Eighth IEEE International Conference on Data
  Mining}, \bibinfo{organization}{IEEE}. pp. \bibinfo{pages}{502--511}.
\bibitem[{Sarwar et~al.(2001)Sarwar, Karypis, Konstan and
  Riedl}]{sarwar2001item}
\bibinfo{author}{Sarwar, B.}, \bibinfo{author}{Karypis, G.},
  \bibinfo{author}{Konstan, J.}, \bibinfo{author}{Riedl, J.},
  \bibinfo{year}{2001}.
\newblock \bibinfo{title}{Item-based collaborative filtering recommendation
  algorithms}, in: \bibinfo{booktitle}{Proceedings of the 10th international
  conference on World Wide Web}, pp. \bibinfo{pages}{285--295}.
\bibitem[{Shao et~al.(2021)Shao, Li and Bian}]{shao2021survey}
\bibinfo{author}{Shao, B.}, \bibinfo{author}{Li, X.}, \bibinfo{author}{Bian,
  G.}, \bibinfo{year}{2021}.
\newblock \bibinfo{title}{A survey of research hotspots and frontier trends of
  recommendation systems from the perspective of knowledge graph}.
\newblock \bibinfo{journal}{Expert Systems with Applications}
  \bibinfo{volume}{165}, \bibinfo{pages}{113764}.
\bibitem[{Sun et~al.(2019)Sun, Liu, Wu, Pei, Lin, Ou and
  Jiang}]{sun2019bert4rec}
\bibinfo{author}{Sun, F.}, \bibinfo{author}{Liu, J.}, \bibinfo{author}{Wu, J.},
  \bibinfo{author}{Pei, C.}, \bibinfo{author}{Lin, X.}, \bibinfo{author}{Ou,
  W.}, \bibinfo{author}{Jiang, P.}, \bibinfo{year}{2019}.
\newblock \bibinfo{title}{Bert4rec: Sequential recommendation with
  bidirectional encoder representations from transformer}, in:
  \bibinfo{booktitle}{Proceedings of the 28th ACM international conference on
  information and knowledge management}, pp. \bibinfo{pages}{1441--1450}.
\bibitem[{Tang and Wang(2018)}]{tang2018personalized}
\bibinfo{author}{Tang, J.}, \bibinfo{author}{Wang, K.}, \bibinfo{year}{2018}.
\newblock \bibinfo{title}{Personalized top-n sequential recommendation via
  convolutional sequence embedding}, in: \bibinfo{booktitle}{Proceedings of the
  eleventh ACM international conference on web search and data mining}, pp.
  \bibinfo{pages}{565--573}.
\bibitem[{Vaswani et~al.(2017)Vaswani, Shazeer, Parmar, Uszkoreit, Jones,
  Gomez, Kaiser and Polosukhin}]{vaswani2017attention}
\bibinfo{author}{Vaswani, A.}, \bibinfo{author}{Shazeer, N.},
  \bibinfo{author}{Parmar, N.}, \bibinfo{author}{Uszkoreit, J.},
  \bibinfo{author}{Jones, L.}, \bibinfo{author}{Gomez, A.N.},
  \bibinfo{author}{Kaiser, {\L}.}, \bibinfo{author}{Polosukhin, I.},
  \bibinfo{year}{2017}.
\newblock \bibinfo{title}{Attention is all you need}.
\newblock \bibinfo{journal}{Advances in neural information processing systems}
  \bibinfo{volume}{30}.
\bibitem[{Walek and Fojtik(2020)}]{walek2020hybrid}
\bibinfo{author}{Walek, B.}, \bibinfo{author}{Fojtik, V.},
  \bibinfo{year}{2020}.
\newblock \bibinfo{title}{A hybrid recommender system for recommending relevant
  movies using an expert system}.
\newblock \bibinfo{journal}{Expert Systems with Applications}
  \bibinfo{volume}{158}, \bibinfo{pages}{113452}.
\bibitem[{Wei et~al.(2019)Wei, Wang, Nie, He, Hong and Chua}]{wei2019mmgcn}
\bibinfo{author}{Wei, Y.}, \bibinfo{author}{Wang, X.}, \bibinfo{author}{Nie,
  L.}, \bibinfo{author}{He, X.}, \bibinfo{author}{Hong, R.},
  \bibinfo{author}{Chua, T.S.}, \bibinfo{year}{2019}.
\newblock \bibinfo{title}{Mmgcn: Multi-modal graph convolution network for
  personalized recommendation of micro-video}, in:
  \bibinfo{booktitle}{Proceedings of the 27th ACM International Conference on
  Multimedia}, pp. \bibinfo{pages}{1437--1445}.
\bibitem[{Xie et~al.(2022)Xie, Zhou and Kim}]{xie2022decoupled}
\bibinfo{author}{Xie, Y.}, \bibinfo{author}{Zhou, P.}, \bibinfo{author}{Kim,
  S.}, \bibinfo{year}{2022}.
\newblock \bibinfo{title}{Decoupled side information fusion for sequential
  recommendation}.
\newblock \bibinfo{journal}{arXiv preprint arXiv:2204.11046} .
\bibitem[{Yan and Zhang(2019)}]{yan2019merging}
\bibinfo{author}{Yan, C.}, \bibinfo{author}{Zhang, Q.}, \bibinfo{year}{2019}.
\newblock \bibinfo{title}{Merging visual features and temporal dynamics in
  sequential recommendation}.
\newblock \bibinfo{journal}{Neurocomputing} \bibinfo{volume}{362},
  \bibinfo{pages}{11--18}.
\bibitem[{Zhang et~al.(2019)Zhang, Zhao, Liu, Sheng, Xu, Wang, Liu and
  Zhou}]{zhang2019feature}
\bibinfo{author}{Zhang, T.}, \bibinfo{author}{Zhao, P.}, \bibinfo{author}{Liu,
  Y.}, \bibinfo{author}{Sheng, V.S.}, \bibinfo{author}{Xu, J.},
  \bibinfo{author}{Wang, D.}, \bibinfo{author}{Liu, G.}, \bibinfo{author}{Zhou,
  X.}, \bibinfo{year}{2019}.
\newblock \bibinfo{title}{Feature-level deeper self-attention network for
  sequential recommendation.}, in: \bibinfo{booktitle}{IJCAI}, pp.
  \bibinfo{pages}{4320--4326}.
\bibitem[{Zhao et~al.(2021)Zhao, Mu, Hou, Lin, Chen, Pan, Li, Lu, Wang, Tian
  et~al.}]{zhao2021recbole}
\bibinfo{author}{Zhao, W.X.}, \bibinfo{author}{Mu, S.}, \bibinfo{author}{Hou,
  Y.}, \bibinfo{author}{Lin, Z.}, \bibinfo{author}{Chen, Y.},
  \bibinfo{author}{Pan, X.}, \bibinfo{author}{Li, K.}, \bibinfo{author}{Lu,
  Y.}, \bibinfo{author}{Wang, H.}, \bibinfo{author}{Tian, C.}, et~al.,
  \bibinfo{year}{2021}.
\newblock \bibinfo{title}{Recbole: Towards a unified, comprehensive and
  efficient framework for recommendation algorithms}, in:
  \bibinfo{booktitle}{Proceedings of the 30th ACM International Conference on
  Information \& Knowledge Management}, pp. \bibinfo{pages}{4653--4664}.
\bibitem[{Zhou et~al.(2020)Zhou, Wang, Zhao, Zhu, Wang, Zhang, Wang and
  Wen}]{zhou2020s3}
\bibinfo{author}{Zhou, K.}, \bibinfo{author}{Wang, H.}, \bibinfo{author}{Zhao,
  W.X.}, \bibinfo{author}{Zhu, Y.}, \bibinfo{author}{Wang, S.},
  \bibinfo{author}{Zhang, F.}, \bibinfo{author}{Wang, Z.},
  \bibinfo{author}{Wen, J.R.}, \bibinfo{year}{2020}.
\newblock \bibinfo{title}{S3-rec: Self-supervised learning for sequential
  recommendation with mutual information maximization}, in:
  \bibinfo{booktitle}{Proceedings of the 29th ACM International Conference on
  Information \& Knowledge Management}, pp. \bibinfo{pages}{1893--1902}.

\end{thebibliography}

\end{document}